\documentclass[onecolumn,showpacs,aps,prl,preprintnumbers,amsmath,amssymb]{revtex4-1}
\usepackage[english]{babel}
\usepackage{amssymb}
\usepackage{graphicx}

\newcommand{\const}{{\rm const}}

\begin{document}
\selectlanguage{english}
\title{Diffusion of large particles through small pores:\\
  from entropic to enthalpic transport}
\author{I.V.Bodrenko, S.Salis, S.Acosta-Gutierrez, M.Ceccarelli\\
{\small \it Department of Physics, University of Cagliari, } \\
{\small \it S.P. Monserrato-Sestu km 0.700, I-09042 Monserrato (CA), Italy}}
%
\begin{abstract}
We present an implicit statistical model for the steric effect on the potential of mean force (PMF)
of a molecule diffusing through a flexible nanochannel of varying size.
The average cross sectional area profile of the channel and the average minimal
projection area of the molecule are the two major quantities determining the
steric part of the PMF barrier for the translocation of the particle
in the case of a small rigid particle and a large rigid channel. 
In this case, the description is reduced to the Fick-Jacobs model and the PMF is completely entropic.
However, the flexibility of channel's cross section and that of molecule's size
play crucial role when a large molecule goes through a narrow channel.
The PMF profile changes its statictical nature and becomes enthalpic.
We treat the flexibility in terms of the equilibrium fluctuations of the pore
and of the molecule, independently. For the case of gaussian fluctuations, we derived simple
analytical expressions for the steric barrier.
\end{abstract}
\maketitle
 
\section{Introduction}
Passive transport of molecules through nanosized channels in porous media and membranes
is of fundamental importance for applications in material science (e.g., particle separation and
filtering \cite{Koros2017}), in nanotechnology (e.g., Brownian motors \cite{Hanggi2009},
molecular sensing \cite{Gu:1999eo, Howorka:2012he})
and in biology (e.g., function of cellular membranes \cite{Cooper1988, Bressloff2013}).

This process may often be adequately described within the diffusion approximation by treating the
molecules as over-damped Brownian particles in the pore \cite{Kramers1940,Hanggi1990,Gardiner2004}.
In general, this assumes the adiabatic separation of the time scales, i.e., the fast degrees of freedom 
are in the thermodynamic equilibrium while few slow ones are evolving quasi-stationary according to a 
stochastic Markov process. 
In many cases, it is sufficient to consider only one coordinate, e.g, the one characterizing
the position of the molecule along the pore or the reaction path. 
The corresponding 1-dimensional Fokker-Planck equation (FPE) for the probability density 
of molecule's coordinate, $\phi(x,t)$, is reduced to the Smoluchowski diffusion-drift equation, 
\begin{eqnarray}
\frac{\partial \phi(x,t)}{\partial t} & = & \frac{\partial}{\partial x} D(x)
    \left(\frac{\partial \phi(x,t)}{\partial x} + \frac{\phi(x,t)}{kT}\frac{\partial U(x)}{\partial x}\right),
                                          \label{FP}
\end{eqnarray}
where $k$ stands for Boltzmann's constant, $T$ is the temperature and $D(x)$ is the position-dependent 
diffusion coefficient.  $U(x)$ is called the potential of mean force (PMF) as it generates the mean drift force
acting onto the Brownian particle, $f(x) = - dU(x)/dx$. The difference between the PMF values at two points,
$x$ and $x=0$, equals the minimal thermodynamic (reversible) work required for moving the molecule
from point $x=0$ to $x$, $U(x)=\int_{0}^{x}f(x')dx' = R_{\rm min}(x)$; here, we set $U(0)=0$.
According to the fundamental principles of thermodynamics \cite{LL5}, the minimal work required to change
a parameter of a system while keeping it in the thermodynamic equilibrium, e.g.,
maintaining constant pressure, $p$, and temperature $T$, is equal to the corresponding difference of the appropriate
thermodynamic potential of the total system (particle plus medium), e.g., the Gibbs free energy
$R_{\rm min}(x)=G(x,p,T)-G(x=0,p,T) \equiv \Delta G(x)$. Thus, 
\begin{eqnarray}
U(x) & = & \Delta G(x). \label{PMF_dG}  
\end{eqnarray}
Equation (\ref{FP}) has a Boltzmann-type equilibrium (zero-flux) solution,
\begin{eqnarray} 
\phi_{\rm eq}(x) = c_0 \exp\left(-U(x)/kT\right),   \label{phi_eq}
\end{eqnarray}
where $c_0$ is a constant dependent on the boundary conditions. The latter equation also
represents the condition of a constant ($x$-independent) equilibrium chemical potential
for the particle \cite{LL5}, and give rise to an alternative interpretation of the
PMF, i.e.,
\begin{eqnarray}
  \mu & = & kT {\rm ln}\left(\frac{\phi_{\rm eq}(x)}{c_0}\right) + U(x) + \psi = \const .
\end{eqnarray}
Here, quantity $\psi=\psi(p,T)$ does not depend on $x$. 

The central input quantities of the model,  $U(x)$ and $D(x)$, may be determined microscopically 
by using individual trajectories from all-atom simulations including, e.g., the pore, the membrane,
the solvent and the molecules, see, e.g., \cite{Im2002,Hummer2005,Wilson2014,Berezhkovskii2017}.
For bigger  molecules and stronger interaction, the timescale for the particle to cross the pore increases
and plain all-atom simulations may become unfeasible to achieve statistically convergent results.
Then, the enhanced sampling methods (e.g., the umbrella sampling \cite{Torrie1974} or the metadynamics
\cite{Laio2002,Tiwary2013}) may be utilized to obtain both the free energy profile and the kinetic
parameters. But even when using the enhanced sampling, the all-atom approach is still often
computationally demanding as multi-microseconds long trajectories may be required
\cite{Hajjar2010,DAgostino:2016bpb,Bajaj2017,Ghai2017}.  

An alternative and complementary approach consists of using an implicit or continuum
physical model for the free energy and the diffusion constant profiles.
The possibilities to treat wide range of the timescales and to obtain the results in an 
analytic form are the major advantages of the implicit method; a limited accuracy is the payoff.

The Poisson-Nernst-Planck (PNP) approach to the transport of atomic ions
constitute a class of implicit models. There, equation (\ref{FP}) is coupled to the Poisson-Boltzmann
model (linearized or non-linear) to describe the ion-ion and ion-medium electrostatic interactions,
the pore size and ion's finite radii are taken into account within the hard spheres / hard  wall model or
by assuming repulsive, $\propto 1/r^{12}$, Van der Waals interaction;
see, e.g., \cite{Levitt1991,Levitt1991II,Horng2012} and the recent review, \cite{Maffeo2012}
for further details.

In a different implicit model, the transport of small particles in a channel of varying radius 
is reduced to the effective one-dimensional diffusion along the channel axis governed by
the Fick-Jacobs equation of type (\ref{FP}), \cite{Zwanzig1992}. Here, the PMF is obtained
by using Eq. (\ref{phi_eq}) and noting that in equilibrium the probability density is
proportional to the cross sectional area of the pore, $\phi_{\rm eq}(x)  \propto  A(x)$ ; and therefore 
\begin{eqnarray}
U(x)-U(0) & = & -kT \ln \left(\frac{A(x)}{A(0)}\right). \label{steric_fes}
\end{eqnarray}
The PMF profile is governed by the channel cross section profile. It was also found, that
the actual validity of the Fick-Jacobs equation goes far beyond the formal adiabatic separation of the time scales
if one introduces a proper position-dependent diffusion coefficient, $D(x)$, which takes into account the 
relaxation kinetics in the transverse coordinates, \cite{Zwanzig1992,Reguera2001,Kalinay2006,Bradley2009}.
This model has inspired several concepts in diffusive transport theory and applications, e.g.,
the entropic transport \cite{Reguera2006}, the entropic filter \cite{Reguera2012,Lairez2016},
the entropic stochastic resonance \cite{Burada2008}. 

In the present work, we consider diffusion of a particle through  a channel with varying cross section and having 
a narrow constriction region, so that the average equilibrium size of the molecule may be larger than the average 
size of the pore. But since the pore can be expanded and the molecule can be compressed, the latter still can be 
translocated over the constriction barrier. To our knowledge, there are no existing implicit models to tackle the
problem without performing all-atom simulations of the transport process of the whole system -- the particle,
the channel and the medium. Our aim here is to connect analytically the PMF of the particle in the pore with
their geometric and flexibility parameters.  

\section{Theory}

\begin{figure}[th]
\begin{center}
\includegraphics[scale=0.40]{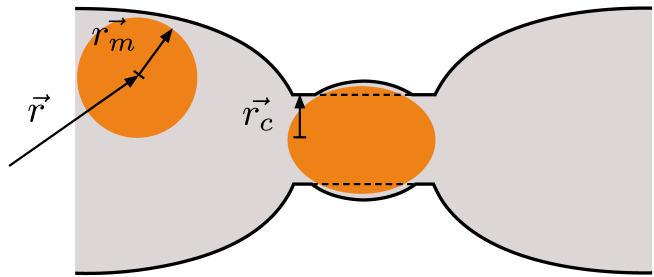}         
\caption{Schematic diagram of a flexible particle diffusing through circular flexible channel. $x$-coordinate
  of vector $\vec{\rm r}$ is along the channel axis. $\vec{\rm r}$ is the smaller radius of particles spheroid,
  and the channel radius $\vec{\rm c}(x)$ at the location of the particle.
\label{fig1}}
\end{center}
\end{figure}

We consider a channel having circular cross section and a spherical particle which may be compressed to a
spheroid with the long axis always along the channel (Fig.\ref{fig1}). Further, we set up a minimal set of
variables characterising the configuration of the particle in the channel, i.e., $\{x,y,z\}$ are the
Cartesian coordinates of the center of the particle ($x$ is along the channel axis); $r_{\rm c}$ is the
radius of the channel at the position of the particle; $r_{\rm m}$ is the smaller radius of the particle spheroid.
The minimal work, $R_{\rm min}=R_{\rm min}(x,y,z,r_{\rm c},r_{\rm m})$,  required to move the particle from a certain point
$\{x_0,y_0,z_0\}$ (e.g., with $x_0=y_0=z_0=0$) and equilibrium values of pore's and particle's radii
to point $\{x,y,z\}$ with radii $r_{\rm c},r_{\rm m}$ determines the probability density for the parameters \cite{LL5},
$\psi(x,y,z,r_{\rm c},r_{\rm m}) \propto \exp \left(-R_{\rm min}/kT\right)$. Then, the equilibrium probability density
for $x$ reads,
\begin{eqnarray}
  \phi_{\rm eq}(x) & \propto& \int \exp\left(-\frac{R_{\rm min}(x,y,z,r_{\rm c},r_{\rm m})}{kT}\right) dy dz ds_{\rm c} ds_{\rm m}.
             \nonumber \\  \label{phi_eq_2}
\end{eqnarray}
Assuming constant $p$ and $T$, the minimal work may be formally calculated as the free energy change along
the thermodynamic cycle,
\begin{eqnarray}
  R_{\rm min}(x,y,z,r_{\rm c},r_{\rm m}) & = & \Delta G_{\rm c}(r_{\rm c},x) + \Delta G_{\rm m}(r_{\rm m}) + \nonumber \\
 &&  \Delta G_{\rm int}(x,y,z,r_{\rm c},r_{\rm m}) . \label{Rmin}
\end{eqnarray}
The first term stands for the contribution due to the change of free pore's radius at point $x$ from its
equilibrium value to $r_{\rm c}$; it determines the equilibrium probability density of pore radius
fluctuations, $f_{\rm c}(r_{\rm c},x)~\propto~\exp \left(-\Delta G_{\rm c}/kT\right)$.
The second term is the contribution from the change of particle's smaller
radius from its equilibrium values to $r_{\rm m}$ outside the pore, and its probability density for 
the free particle reads, $f_{\rm m}(r_{\rm m}) \propto \exp \left(-\Delta G_{\rm m}/kT\right)$.
The last term in (\ref{Rmin}) is the interaction contribution to the minimal work.
It contains all the effects of the particle pore interaction, and it vanishes by definition when the particle coordinates,
$\{x,y,z\}$, are outside the pore.
Here, we focus on the steric contribution to the interaction free energy which originates from the fact that
the atoms have finite size and can not overlap. Assuming the hard-wall repulsion between
the channel and the particle, we define the steric interaction potential as follows, -- it equals zero when the channel
and the particle do not overlap; it it positively infinite otherwise. For the coordinate system chosen in Fig.\ref{fig1},
one obtains
\begin{eqnarray}
  G_{\rm int}(x,y,z,r_{\rm c},r_{\rm m}) & = & \left\{
  \begin{array}{cc}
    0       &, \sqrt{y^2+z^2}+r_{\rm m} < r_{\rm c}(x) \\
    +\infty &, \sqrt{y^2+z^2}+r_{\rm m} > r_{\rm c}(x) 
  \end{array}  \right. \nonumber \\ \label{Gsteric}
\end{eqnarray}
By collecting the above results between Eqs.(\ref{phi_eq_2}) and (\ref{Gsteric}), one obtains,
\begin{eqnarray}
 \phi_{\rm eq}(x) & \propto & A(x),
\end{eqnarray}
where
\begin{eqnarray}
  A(x) & \equiv & \pi \int_0^\infty t^2  f_{\rm cm}(t,x)  dt \label{defA}
\end{eqnarray}
has the meaning of the average available cross sectional area for the particle inside the pore at coordinate $x$,
and 
\begin{eqnarray}
  f_{\rm cm}(t,x) & \equiv & \int f_{\rm c}(r_{\rm m}+t,x) f_{\rm m}(r_{\rm m}) d r_{\rm m}. \label{def_f_cm}
\end{eqnarray}
may be formally interpreted as the equilibrium probability density for the particle to be centered at distance $t$
from the channel center when it is located at $x$ along the channel axis. Finally, by taking into account (\ref{phi_eq}),
the PMF may be calculated with Eq.(\ref{steric_fes}), but assuming Eq.(\ref{defA}) for $A(x)$.
Equations (\ref{steric_fes}) and (\ref{defA},\ref{def_f_cm}) give the steric contribution to the potential of mean force
of a molecules diffusing in the channel taking into account both their average size and the flexibility.
In the limit of a point-like particle and hard-walled channel, $A(x)$ becomes channel's cross sectional area, and
Eq. (\ref{steric_fes}) is reduced to the Fick-Jacobs approximation..
The input quantities, i.e., probability densities $f_{\rm c}(r_{\rm c},x)$ and $f_{\rm m}(r_{\rm m})$, may be calculated
by studying the equilibrium fluctuations of $r_c(x)$ and of $s_m$
in  all-atom MD simulation, separately, for the channel and for the molecule. As the fluctuation are not necessarily
small, the advanced sampling techniques may be used to accelerate the statistical convergence.

The major assumption behind the present model consists in using the hard-wall repulsion model,
(\ref{Gsteric}), for the interaction free energy. This may be argued by recalling the well-known
fact that the potential energy is stored preferably in the soft degrees of freedom under the mechanical
equilibrium. As the Van der Waals inter-atomic repulsion is very stiff ($\propto 1/r^{12}$) compared
to the potential energy function of molecular bonds, angles and dihedrals responsible for the channel/pore
flexibility, the external work stored (in the form of free energy) in the channel-particle inter-atomic
contacts is much smaller than $\Delta G_{\rm c}$ and $\Delta G_{\rm m}$. Therefore, the major effect of the steric
interaction consists in the limitation of the configuration space.

The selection of the collective variables characterizing the channel/particle deformations
is another important approximation of the model. We have utilized the minimal set of variables
in our example (see, Fig.\ref{fig1}), thus neglecting, e.g., the free energy related to the possible rotation of
the long axis of particle's spheroid with respect to the axis of diffusion.

As far as the PMF is used to characterize the thermodynamic equilibrium, (\ref{phi_eq}), there is no question
of the timescales. The story changes when one uses the PMF in the diffusion transport equation, (\ref{FP}).
Then, one should suppose the adiabatic separation of $x$ from all other degrees of freedom
(including those used to calculate $U$) assuming the thermodynamic equilibrium for them. Fortunately,
this very strong condition may be soften and the validity of Eq.(\ref{FP}) may be extended if
one introduces appropriate position-dependent diffusion coefficient, \cite{Hanggi1990,Berezhkovskii2011}.

To illustrate better the obtained results, we consider in more detail a natural and important case
when $f_{\rm c}(r_{\rm c},x)$ and $f_{\rm m}(r_{\rm m})$ are normal distributions with the average values,
correspondingly, $R_{\rm c}(x)$ and $R_{\rm m}$, and the dispersions $\sigma_{\rm c}(x)$ and $\sigma_{\rm r}$,
respectively. Distribution $f_{\rm cm}(t,x)$ is also a Gaussian having the mean value and the variance,
$R_{\rm cm}(x)=R_{\rm c}(x)-R_{\rm c}$ and $\sigma^2_{\rm cm}(x) = \sigma^2_{\rm c}(x)+\sigma^2_{\rm cm}$, respectively.
The average available area defined by Eq. (\ref{defA}) reads,
\begin{eqnarray}
A(x) & = & \pi \int_0^\infty t^2 
   \frac{\exp\left(-\frac{(t-R_{\rm cm}(x))^2}{2 \sigma^2_{\rm cm}(x)}\right)}{\sqrt{2 \pi\sigma^2_{\rm cm}(x) }} d t .
                     \label{A_fluct_gauss}
\end{eqnarray}
Obviously, the integral in the latter equation may be solved analytically in terms of the ${\rm erf}$ function
and the exponential function, but the integral form, (\ref{A_fluct_gauss})
is more compact and useful for the following analysis.

If the channel is wider than the molecule, $R_{\rm cm}(x) > 0$, and both are stiff,
$R_{\rm cm}(x)/\sigma_{\rm cm}(x) \gg 1$, then the Gaussian in (\ref{A_fluct_gauss}) may be approximated by the
Dirac delta function and one obtains,
\begin{eqnarray}
  A(x) & \approx & \pi R^2_{\rm cm}(x) . \label{A_gauss_1}
\end{eqnarray}
I.e., the PMF profile is determined by the difference of the pore and the particle equilibrium radii,
in agreement with the extensions of the Fick-Jacobs equation to finite-sized particles, \cite{Reguera2012}.

If the particle barely fit the pore at some region, so that $|R_{\rm cm}(x)/\sigma_{\rm cm}(x)| \ll 1$, then
\begin{eqnarray}
  A(x) & \approx & \pi \sigma^2_{\rm cm}(x)/2 . \label{A_gauss_2} 
\end{eqnarray}
Therefore the PMF has a different physical origin and is completely determined by the
pore/particle flexibility in terms of the equilibrium fluctuations.

Finally, in the limit of a narrow channel and a thick molecule,
$R_{\rm cm}(x) < 0$ and $|R_{\rm cm}(x)|/\sigma_{\rm cm}(x) \gg 1$, 
one may limit integration in (\ref{A_fluct_gauss}) 
around $|R_{\rm cm}(x)|~\approx~0$ in the small interval of the order on magnitude of
$\sigma^2_{\rm cm}(x)/|R_{\rm cm}(x)|$ where
\begin{eqnarray}
\exp\left(-\frac{(t-R_{\rm cm}(x))^2}{2 \sigma^2_{\rm cm}(x)}\right) & \approx & 
           \exp\left(-\frac{t |R_{\rm cm}(x)|}{\sigma^2_{\rm cm}(x)} 
              -\frac{R^2_{\rm cm}(x)}{2 \sigma^2_{\rm cm}(x)}\right). \nonumber
\end{eqnarray}
Then, one arrives at the following estimate
\begin{eqnarray}
A(x) & \approx & \sqrt{\frac{2}{\pi}} \frac{\sigma^5_{\rm cm}(x)}{|R_{\rm cm}(x)|^3} 
     \exp\left(-\frac{R^2_{\rm cm}(x)}{2 \sigma^2_{\rm cm}(x)}\right).
              \label{A_gauss_3}
\end{eqnarray}
Therefore, in the case of a bulky molecule diffusing through a tight channel,
the average available area is exponentially small, and depends on both the average radii
and on the fluctuations. By inserting (\ref{A_gauss_3}) into (\ref{steric_fes}) and
neglecting the logarithmic terms of $|R_{\rm cm}(x)|/\sigma_{\rm cm}$ compared with the
linear ones, one obtains the estimate of the steric barrier with the logarithmic accuracy,
\begin{eqnarray}
 U(x)-U(0) & \approx & kT \frac{R^2_{\rm cm}(x)}{2\sigma^2_{\rm cm}(x)} \label{StericBarrier}. 
\end{eqnarray}
The latter equation also shows that, in contrast with the Fick-Jacobs limit of a small particle,
(\ref{A_gauss_1}), when the PMF profile has logarithmic dependence on the pore/molecule size,
the opposite limit demonstrates much stronger power-low dependence on  $|R_{\rm cm}(x)|/\sigma_{\rm cm}$.

From the above analysis, one may also conclude that the PMF changes its statistical origin with the
relative pore/molecule size from purely entropic in the Fick-Jacobs limit \cite{Zwanzig1992}
to essentially enthalpic in the opposite case of a large molecule. Indeed, as $R_{\rm cm}(x)$
may be considered independent or weakly dependent on the temperature, the PMF in
the Fick-Jacobs limit is proportional to temperature, and the entropy
$\Delta S(x) = \partial U/\partial T \approx  -k \ln \big( R^2_{\rm cm}(x)/R^2_{\rm cm}(0) \big)$.
In contrast, the fluctuations of the molecular size, according the fluctuation-dissipation theorem,
\cite{LL5}, are proportional to temperature, $\sigma^2_{\rm m} = \alpha T $;
here $\alpha=|dR_{\rm m}/df|$ determines the compressibility
of the molecule under external force $f$ in the limit $f\rightarrow 0$.
Analogously, we can conclude for the channel radius.
If $\alpha$ does not depend (or depend weakly) on the temperature,
the steric barrier (\ref{StericBarrier}) becomes temperature-independent, i.e.,
has enthalpic properties, as the entropy, $S(x) = \partial U/\partial T \approx  0$.

\section{Conclusion}

To summarize, we have presented an implicit model describing the contribution of the
steric constraint to the potential of mean force of a molecule diffusing through a nano-sized channel.
The statistical nature of the PMF changes with the relative pore/molecule radius.
If the channel is much wider than the molecule, the model is reduced to the Fick-Jacobs approximation,
describing the PMF in terms of the average radii of the channel and the molecule, and the PMF is
purely entropic. In the opposite case of a large molecule and a small pore the PMF also depends on the
flexibilities of the two, and it is essentially enthalpic.
The channel/pore flexibility is described in terms of the equilibrium
fluctuations of their size and may be calculated by using all-atom simulations of the pore and of the
molecule, independently, what may significantly reduce the computational complexity.

Although the model is presented here in a simplistic form (we assume one-dimensional diffusion,
circular channel, spheroidal particle, etc.),  the suggested way of reasoning allows one, in principle, 
to make a straightforward extension to the case of several collective variables,
$x=\{x_1,x_2,\ldots\}$ and to take into account other interaction terms
(e.g., electrostatic, hydrophobic, etc.).

The presented method may be especially useful for the problems of diffusive transport of
a bulky molecule trough a small flexible pore where the effect of steric constraints creates a significant
barrier for the passing molecules.
For example, this is the case when a nutrient or an antibiotic molecule passes through a porin \cite{Pages2008}.
In another example, the uptake of hydrophobic molecules into the cellular membrane takes
place via the spontaneous opening in the lateral part of a membrane protein, \cite{Hearn2009}.
Discovered here strong dependence of the steric barrier on the particle size and flexibility,
see Eq.(\ref{A_gauss_2}), could potentially be utilized in new designs of molecular filters.

\section{Acknowledgment}
The research leading to these results was conducted as part of
the Translocation consortium \\ (http://www.translocation.com)
and received support from the Innovative Medicines Initiatives
Joint Undertaking under grant agreement no. 115525,
resources which are composed of financial contribution from
the European Union's seventh framework programme (FP7/
 2007-2013) and EFPIA companies' kind contribution.
 
\bibliography{steric_fes} 
\bibliographystyle{apsrev4-1}

\end{document}